\documentclass[conference]{IEEEtran}
\IEEEoverridecommandlockouts

\usepackage[utf8]{inputenc}
\usepackage[T1]{fontenc}

\usepackage{graphicx}
\usepackage{physics}
\usepackage{amsmath}
\usepackage{amssymb}
\usepackage{xcolor}
\usepackage{comment}
\usepackage{hyperref}

\begin{document}

\title{Network-based link prediction of scientific concepts -- a Science4Cast competition entry.}

\author{
\IEEEauthorblockN{João P. Moutinho}
\IEEEauthorblockA{\textit{Physics of Information and}\\
\textit{Quantum Technologies Group} \\
\textit{Instituto de Telecomunicações}\\
Lisbon, Portugal \\
joao.p.moutinho@gmail.com}
\and
\IEEEauthorblockN{Bruno Coutinho}
\IEEEauthorblockA{\textit{Physics of Information and}\\
\textit{Quantum Technologies Group} \\
\textit{Instituto de Telecomunicações}\\
Lisbon, Portugal \\
bruno.coutinho@lx.it.pt}
\and
\IEEEauthorblockN{Lorenzo Buffoni}
\IEEEauthorblockA{\textit{Physics of Information and}\\
\textit{Quantum Technologies Group} \\
\textit{Instituto de Telecomunicações}\\
Lisbon, Portugal \\
lorenzobuffoni@gmail.com}
}

\maketitle

\begin{abstract}

We report on a model built to predict links in a complex network of scientific concepts, in the context of the Science4Cast 2021 competition. We show that the network heavily favours linking nodes of high degree, indicating that new scientific connections are primarily made between popular concepts, which constitutes the main feature of our model. Besides this notion of popularity, we use a measure of similarity between nodes quantified by a normalized count of their common neighbours to improve the model. Finally, we show that the model can be further improved by considering a time-weighted adjacency matrix with both older and newer links having higher impact in the predictions, representing rooted concepts and state of the art research, respectively.

\end{abstract}


\section{Introduction}

The Science4Cast 2021 competition \cite{Science4Cast, bigdata2021} addresses the challenge of predicting future connections in the ever-growing semantic network of scientific concepts used in the fields of Machine Learning and Artificial Intelligence in order to foresee which research topics will emerge in the coming years. In the competition we are given two instances of this dataset: one corresponding to the state of the network in 2014, and one to 2017. The ultimate goal would be to predict the state of the network in 2020, but the task at hand in the challenge is a simpler one: given a random set composed of $10^6$ links that do not exist in 2017 we must order them according to which we think have the best chances of being connected by 2020. For the 2014 to 2017 predictions we are given a similar random set for which we know the solution in 2017, allowing us to train a prediction model.

The challenge in this competition is closely related to the active field of link prediction in complex networks \cite{Liben:2007, Lu:2011, wang2015link}. Many well-known network-based link prediction methods such as Common Neighbours \cite{Liben:2007}, Resource-Allocation \cite{Zhou:2009} and Adamic-Adar \cite{adamic2003friends} follow one simple principle: two nodes are more likely to connect if they have many common neighbours. This is equivalent to counting paths of length 2, and typically implies similarity between nodes, favouring the creation of triangles in the network ($A\sim B$, $B\sim C$ $\Rightarrow$ $A\sim C$). In the context of predicting protein-protein interactions it has been shown that counting paths of length 3 is in fact a better predictor than paths of length 2 \cite{kovacs2019network}, the insight being that proteins interact not because they are similar between themselves, but because they are similar to each other's neighbours. This principle favours the creation of squares in the network. A more recent result \cite{qlp} shows that a quantum algorithm for link prediction based on continuous-time quantum walks can identify predictions based on both even and odd-length paths, which sometimes improves the precision over the aforementioned methods. The method encodes the prediction scores in the amplitudes of a quantum superposition, allowing the best links to be sampled from the distribution, avoiding the need to explicitly calculate all pair-wise scores in the network. When ran on a quantum computer, this method can potentially provide a speedup in link prediction.

\subsection{Technical details}
In both the 2014 and 2017 versions of the dataset we are given a list of links $(i,j,t_{ij})$ where $i$ and $j$ are two nodes that formed a link at a certain time $t_{ij}$. In our prediction models we mostly deal with the adjacency matrix $A$ of the network, which we build by adding a weight $w_{ij}$ to each entry $A_{ij}$ and $A_{ji}$ for each $(i, j)$ present in the original list. For the unweighted case we simply have $w_{ij} = 1$. We note that the dataset often has repeated links for the same pair of nodes $(i,j)$, which makes sense given that two concepts can be linked more than once at different times in different research articles. As such, even the unweighted case where all $w_{ij} = 1$ leads to an adjacency matrix where the entries $A_{ij}$ can be greater than 1, which is by itself a weighted adjacency matrix. 

We also deal with the degree of the nodes which we denote by $k_i$ for a node $i$, and which can be computed directly from the adjacency matrix,
\begin{equation}
    k_i=\sum_jA_{ij}.
\end{equation}
Finally we deal with common neighbours between nodes. Let $\Gamma(i)$ be the set of nodes neighbouring $i$. The number of common neighbours between $i$ and $j$ can be computed from the second power of the adjacency matrix,
\begin{equation}
    |\Gamma(i)\cap\Gamma(j)| = (A^2)_{ij}.
\end{equation}

\section{Preliminary data analysis}
\subsection{Prediction set degree distribution}

As mentioned, we are given a list of randomly selected pairs of nodes that we must order from best to worst prediction. In both the training data and competition data this list contains $10^6$ pairs, which is only a very small subset of the full list of possible predictions, with a size on the order of $10^9$ for a sparse network with $N\sim 64,000$. In order to gain some insight into the types of models that will work best for this task it is useful to see how the nodes in each potential link that we must score are connected to the network. As such, for each pair $(v_1, v_2)$ we calculated the degree values $(k_1, k_2)$ and identified three categories of potential links: links between two nodes with zero degree, links between one node with zero degree and one with degree greater than zero, and links between two nodes with degree greater than zero. We show in Table \ref{tab:predictiondegrees} the number of pairs that fall in each category for both the 2014 and 2017 networks.

\begin{table}
    \centering
    \caption{Number of pairs of nodes that fall in each of the three degree categories.}
    \label{tab:predictiondegrees}
    \begin{tabular}{ccc}
    \hline\hline
         & Training (2014) & Competition (2017) \\\hline
    $(k = 0, k = 0)$ & 265966 & 32571 \\
    $(k = 0, k > 0)$ & 500060 & 297545\\
    $(k > 0, k > 0)$ & 233974 & 669884\\\hline\hline
    \end{tabular}
\end{table}

From Table \ref{tab:predictiondegrees} we note that there is a large quantity of links to be scored which include nodes with $k=0$, especially in the training set. This indicates that models solely based on paths between nodes, which are common in network-based link prediction, should not perform well in this task since they can only score links belonging to the same connected component of the network. Furthermore, we note that the distribution of pairs between each category is quite different between the training set and the competition set, which may introduce some difficulties in training a model suitable for the competition set.

\subsection{Network degree distribution}

We now look at the degree distribution of the whole network. Real-world complex networks are often called \textit{scale-free}, meaning that their degree distribution follows a power law \cite{barabasi2016network, RevModPhys.80.1275, Newman2010},
\begin{equation}
P(k)\propto k^{-\gamma},
\end{equation}
with $\gamma>2$, and where $P(k)$ is the density function describing the probability of finding a node with degree $k$ in that network. Values in the $2<\gamma<3$ range are common, and it is a region of interest since the second moment of the distribution diverges \cite{barabasi2016network, RevModPhys.80.1275, Newman2010}. Networks with $\gamma\leq 2$ are not well defined given that in such scenario the average degree diverges.
A scale-free distribution can be explained from a preferential attachment process, where new nodes that are added the network tend to connect to other nodes with a probability proportional to their degree,
\begin{equation}
\Pi(k_i)\propto k_i+B,
\end{equation} 
 with $\Pi(k_i)$ being the probability that a new node connects to a node with degree $k_i$, and $B$ is just a constant. As $B$ increases, the degree of the nodes becomes less relevant. $B$ can be related to $\gamma$ as $\gamma=4+B$ \cite{barabasi2016network,RevModPhys.80.1275,Newman2010}. Lower values of $\gamma$ thus mean that new nodes added to the network are more likely to be connected to nodes with higher degrees. Although this model does not consider new links that are created between nodes that are already part of the network, the same principle can be easily generalized to such a scenario.
\begin{figure}
    \centering
    \includegraphics[width=\columnwidth]{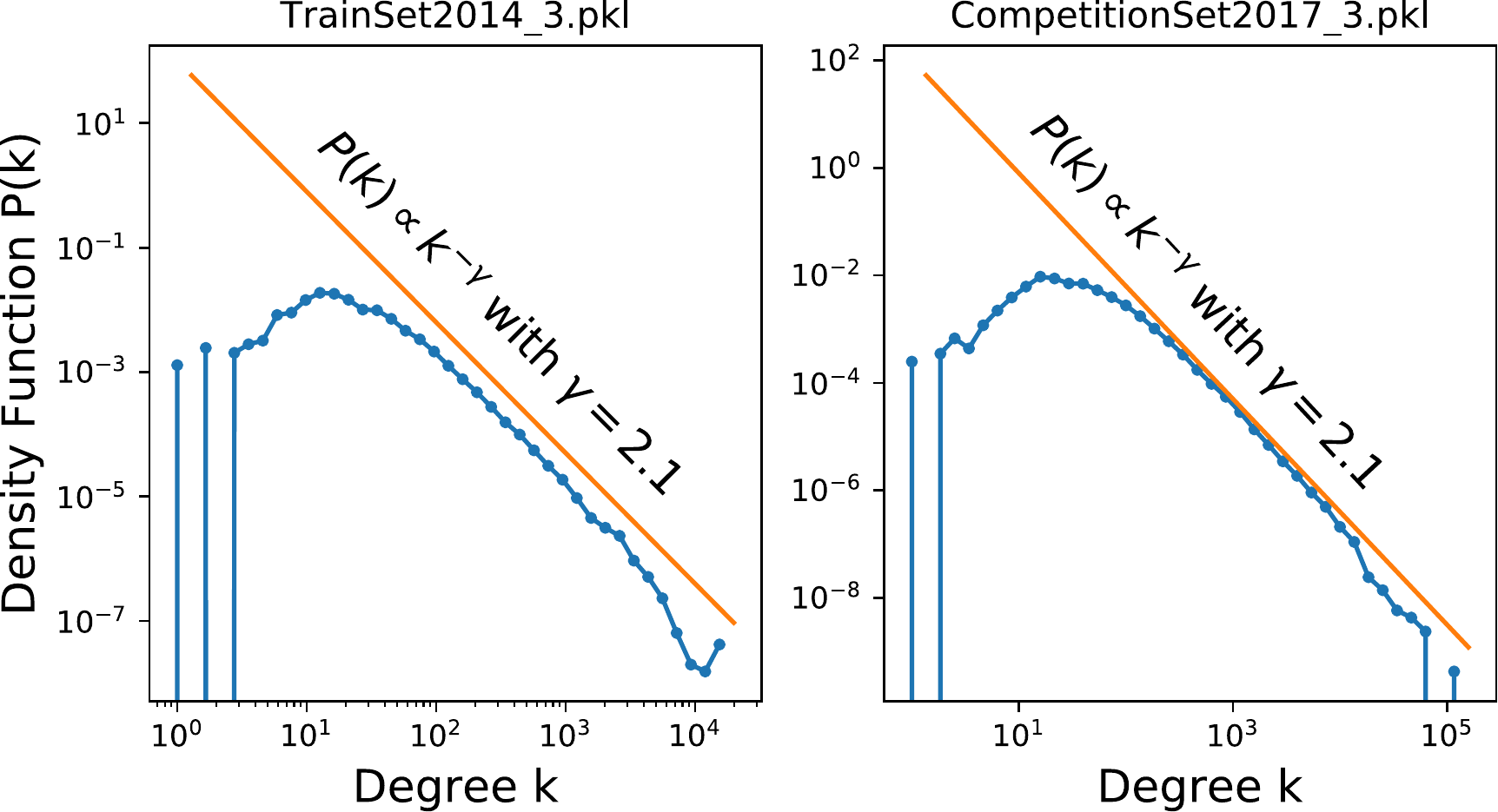}
    \caption{Degree distribution for both the 2014 and 2017 datasets.}
    \label{fig:degreedistribution}
\end{figure}

In Figure \ref{fig:degreedistribution} we plot $P(k)$ for both the 2014 and 2017 networks, and fit a power law to the tail of the distribution. We find that in both cases we get a good fit for $\gamma=2.1$, which is close to the lower bound typically found in complex networks. This indicates that this dataset is likely to grow via a preferential attachment mechanism, and that will be the first hypothesis we will test. Intuitively, given the context of this dataset, this makes sense, as it is reasonable to consider that the popularity of each concept (measured by the degree of the respective node) will play an important role in how new connections are made with that concept.

\section{Network-based methods}

In this section we overview the two main methods we used to predict new connections in this competition. For the results described in this section we disregarded the time-stamp information provided in the data and simply built an adjacency matrix for the 2014 and 2017 networks given by the full list of links in the respective data files.

\subsection{Preferential attachment}

In the previous section we concluded that the growth of this dataset is likely to follow a preferential attachment mechanism. Preferential attachment scores in link prediction are often quantified as
\begin{equation}
    s_{ij}^\text{PA}=k_i*k_j.
\end{equation}
However, this assumes the scoring of links between nodes that are already connected to the network, that is $k_{i,j}>0$, which as we already saw is not the case for all the links we must score in the competition. A possible description for the more general case of scoring both incoming nodes and already connected nodes can be described as 
\begin{equation}
    s_{ij}^\text{PA}=k_i + k_j + \epsilon\sqrt{k_i*k_j}.
\end{equation}
with $\epsilon$ a free parameter. In our tests with the training set we found that $\epsilon=0$ always performed better, and thus we defined our preferential attachment model (PA model) as
\begin{equation}
    s_{ij}^\text{PA}=k_i + k_j. \label{eq:pa}
\end{equation}
Using this simple model we ran our first test on the competition data. Evaluating Eq. \ref{eq:pa} for each pair $(i,j)$ in the set of $10^6$ unconnected pairs we submitted the ordered list to the competition and obtained an AUC value of 0.89715. Immediately we note that PA outperforms the baseline Machine Learning model provided in the competition, an indication that our initial hypothesis was correct.

\subsection{Path-based}

While the preferential attachment model we derived performed well, it uses no information about the distance between $i$ and $j$, which is a popular feature used in link prediction methods, as mentioned in the introduction. As such we decided to test a selection of different path-based methods, including L3 \cite{kovacs2019network}, Common Neighbours \cite{Liben:2007}, Resource Allocation \cite{Zhou:2009} and Adamic-Adar \cite{adamic2003friends}. Ultimately, we found that all of them performed very similarly, indicating that different path structures are not very well defined in the network. Due to time and resource limitations we were unable to test a classical simulation of the quantum walk based method QLP \cite{qlp}, which includes contributions from higher order paths. Nevertheless, given the similar performance between the tested methods, it is unlikely that QLP would have performed better. Although there were only slight differences between the tested methods, we found Adamic-Adar (AA Method) to be the best performing one, with the scores defined in the unweighted case as
\begin{equation}
s_{ij}^\text{AA}=\sum_{u\in\Gamma(i)\cap\Gamma(j)}\frac{1}{\log k_u}
\label{eq:aa}
\end{equation}
where $\Gamma(i)\cap\Gamma(j)$ defines the set of common neighbours between nodes $i$ and $j$. Adamic-Adar is essentially a normalized count of nodes $u$ in this set. The penalization given by $1/\log(k_u)$ increases the importance of common neighbours with lower degree, the intuition being that those neighbours are more unique to $i$ and $j$ compared to other neighbours that may have more connections to other nodes. While Eq. \ref{eq:aa} is written for the unweighted case, in our code we use an adjacency matrix implementation which automatically incorporates the weights in the matrix.

Evaluating Eq. \ref{eq:aa} for each pair $(i,j)$ in the set of $10^6$ unconnected pairs and submitting the ordered list to the competition we obtained an AUC value of 0.87091, which is not as good as the PA method, but is still close to the baseline Machine Learning model.

\subsection{Combining methods}

So far we have described two different models to score links in the dataset: the popularity based PA method, and the similarity based AA method. In order to see if we can improve our overall results we wish to join the results from both methods. To do so we chose a simple linear combination of the normalized score arrays:
\begin{equation}
    \textbf{s}^\text{PA+AA}=\epsilon*\textbf{s}^\text{AA}+(1-\epsilon)*\textbf{s}^\text{PA}
\end{equation}
with $\epsilon$ a free parameter, and where $\textbf{s}^\text{AA}$ and $\textbf{s}^\text{PA}$ are the full arrays of scores for the set of $10^6$ unconnected pairs. Running a grid search for $\epsilon$ between 0 and 1 in the training set we found the best result for $\epsilon=0.92$. Using this value we submitted a combination of the previous AA and PA scores to the competition to obtain an AUC of 0.91385, a small improvement over the standalone PA method. This result indicates that the neighbourhood of the nodes does have some information that impacts the predictions which is not captured by their degree.

\section{Time-weighted adjacency matrix}
In the analysis made so far we purposefully neglected the information that we have about the time $t_{ij}$ at which these links were created. Indeed, the fact that this network isn't a static one but is continuously adding new link is a fact that can't be ignored. Links which are older in time might, for example, represent really well established links between core concepts of the fields, whether newer links might suggest which are the "hot topics" of the research at the moment. In order to include this information in our link prediction methods, we decided to weight the links in the dataset as a function of the time at which they were created. More precisely we considered:
\begin{equation}
    (i, j, t_{ij}) \rightarrow (i, j, f_{\theta}(\tilde{t}_{ij})),
\end{equation}
where $\tilde{t}_{ij}$ is the time $t_{ij}$ at which the link $(i,j)$ was created normalized to a value between 0 and 1, and $f(\cdot)$ is a generic function parameterized by some parameters $\theta$ that we will optimize in order to fit our problem. The purpose here is that by building the adjacency matrix for the network with the weighted links we effectively alter their importance for each method.

\begin{figure}
    \centering
    \includegraphics[width=0.8\columnwidth]{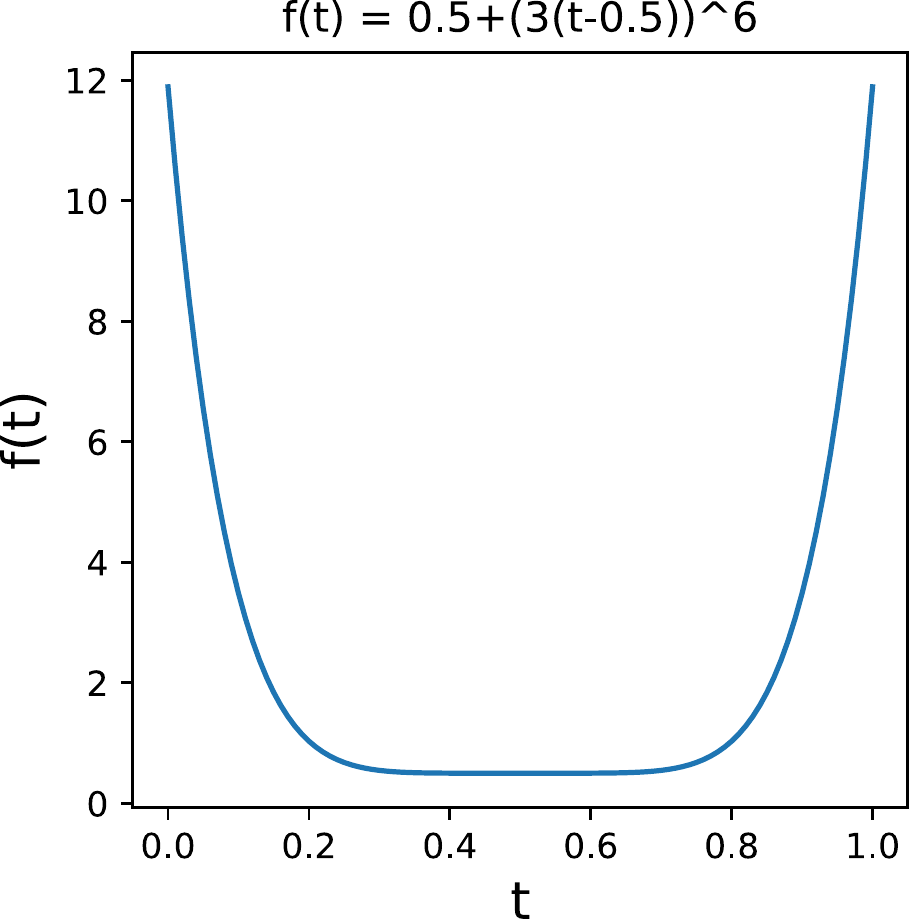}
    \caption{Function used to define the time-weights in the network, where $t$ corresponds to the time-stamp in each link normalized to a value between 0 and 1, and f(t) corresponds to the link weight attributed to that time-stamp.}
    \label{fig:time_function}
\end{figure}

To optimize $f_{\theta}$ one could resort to various techniques that make use of really general function approximators (e.g. Neural Networks). However, given the size of our problem and the computational constraints we started by restricting our optimization with some theoretical assumptions. As we hinted at earlier, our assumptions were exactly that the most important links are both those that are older and newer in time, representing rooted concepts and "hot topics", respectively. These assumptions can be easily quantified by a convex polynomial, and thus we parameterized our time-weighting function as:
\begin{equation}
    f_{\theta}(t) = \theta_0 + (\theta_2\cdot(t-\theta_1))^{\theta_3}
\end{equation}
as illustrated in Figure \ref{fig:time_function} and optimized over the training set using both a general purpose optimizer from scipy and a greedy grid-search while maximizing the AUC for the PA method, given that this is both the more important and more efficient method. In the end the best set of parameters obtained from the training set was $(\theta_0, \theta_1, \theta_2, \theta_3) = (0.0, 0.45, 3, 6)$. We then repeated all the scoring methods described in the previous section, substituting the original adjacency matrix with our time-weighted one, and got an improved result of $\text{AUC}=0.90364$ for the PA method alone, and $0.91385$ for a combination of AA and PA with $a=0.95$. Through some more manual tests on the competition set we ultimately found the best result combining PA and AA of $\text{AUC}=0.91853$ for $(\theta_0, \theta_1, \theta_2, \theta_3) = (0.5, 0.5, 3, 6)$.

It is interesting to see that our hypothesis regarding older and newer links did indeed improve the results. A different optimization routine using a small Neural Network also produced functions showing a similar shape to those obtained from the polynomial model, albeit with reduced performance. Overall though, the improvement we have obtained in this section is quite small, and thus we can not claim this is a strong organizational principle in the growth of the network.

\section{Conclusions}
\begin{table}
    \centering
    \caption{Summary of results and learned parameters. The ranking does not necessarily reflect our preferred Bacalhau dishes.}
    \label{tab:summary}
    \begin{tabular}{c|c|cccc|c|c}
    \hline\hline
    Method & $\epsilon$ & $\theta_0$ & $\theta_1$ & $\theta_2$ & $\theta_3$ & AUC & Bacalhau Code\\\hline
    PA      & -    & - & - & - & - & 0.89715 & à Brás \\
    AA      & -    & - & - & - & - & 0.87091 & à G. de Sá\\
    PA + AA & 0.92 & - & - & - & - & 0.91385 & com Todos\\
    PA      & -    & 0.0 & 0.45 & 3.0 & 6.0 & 0.90364 & com Natas\\
    PA + AA & 0.95 & 0.5 & 0.5 & 3.0 & 6.0  & 0.91853 & à Lagareiro \\\hline\hline
    \end{tabular}
\end{table}

We present in Table \ref{tab:summary} a summary of the results obtained in this competition. Despite it being a simple model composed of handcrafted features based on known organizational principles of complex networks, our scoring method resulted to be competitive with respect to state of the art Machine Learning techniques in the Science4Cast 2021 competition, achieving 4th place in the leaderboard with less than $0.01$ difference in the AUC compared to the top positions.

It is worth mentioning that we did test other more complex models using more features such as predictions from the L3 method, Resource Allocation, Principal Component Analysis, and other node popularity measures such as the Eigenvector Centrality and Average Neighbour Degree. Ultimately, we found that neither of these methods improved our results, even when running a more complex optimization routine combining all of them. Regarding the time-weighting, as we mentioned earlier, we also tried a more general optimization routine using a small Neural Network representation of $f$, but found similar function shapes to those produced by the polynomial model.

Ultimately, the simplicity of our model is also its strongest feature. Even considering just the unweighted preferential attachment model from Eq.\ref{eq:pa}, we were able to reach an AUC close to 0.9 without training any parameters and negligible computational cost. This method can be easily applied to predict new links in the full dataset of 64000 nodes, or even larger datasets that follow the same growth principles. Adding the Adamic-Adar scores also does not require too many resources, consisting essentially of one sparse matrix product and one free parameter to train afterwards for the linear combination of the two score vectors. Optimizing the time-weights is what requires the most resources during training, but impose no added complexity to the models afterwards.

As for a general comment and future directions for the underlying challenge of this competition, we would like to briefly discuss the dataset itself. The dataset provided consists of a semantic network where links represent connections made between concepts in scientific papers. This network is in fact a projection of a larger bipartite network with one group of nodes representing papers and another representing concepts, and links existing solely between papers and concepts. This larger bipartite network is considered a primary network, as it is a more accurate description of the underlying system being modeled. When projecting down to a concept only network, it is likely that some of the organizational principles behind the growth of this system become lost. This type of analysis is described in more detailed in \cite{vasques2018degree}, where, for example, it is shown how the degree distribution of the projected network (as we plotted in Fig. \ref{fig:degreedistribution} for this network) relates to the degree distributions of both node types in the original network. As such, we believe that studying the underlying bipartite system will allow for the creation of better and more precise models to predict future connections. At the same time, the prediction task also becomes harder, as the goal will now be to predict the appearance of new nodes (papers) and the links that come with it (concepts studied). 

\section*{Code Availability}

The code used in this work is available at:\\ \url{https://github.com/Buffoni/quantum-link-prediction}

\section*{Acknowledgements}
The authors thank the support from Funda\c{c}\~{a}o para a Ci\^{e}ncia e a Tecnologia (FCT, Portugal), namely through project UIDB/50008/2020, as well as from projects TheBlinQC and QuantHEP supported by the EU H2020 QuantERA ERA-NET Cofund in Quantum Technologies and by FCT (QuantERA/0001/2017 and QuantERA/0001/2019, respectively), and from the EU H2020 Quantum Flagship project QMiCS (820505). Furthermore, JPM acknowledges the support of FCT through scholarship SFRH/BD/144151/2019 and BC thanks the support from FCT through project CEECINST/00117/2018/CP1495.

\bibliographystyle{IEEEtran}
\bibliography{main}

\end{document}